\documentclass[twocolumn,epsfig,prl,aps]{revtex4} 
\usepackage{psfrag}
\usepackage{graphicx}
\usepackage{dcolumn}
\usepackage{bm}
\usepackage{natbib}
\begin{document}

\title{Hydrodynamic theory of de-wetting}
\author{Jens Eggers}

\affiliation{
School of Mathematics, 
University of Bristol, University Walk, \\
Bristol BS8 1TW, United Kingdom 
        }
\begin{abstract}
A prototypical problem in the study of wetting phenomena
is that of a solid plunging into or being withdrawn 
from a liquid bath. In the latter, de-wetting case, 
a critical speed exists above which a stationary contact line 
is no longer sustainable and a liquid film is being deposited on 
the solid. Demonstrating this behavior to be a hydrodynamic
instability close to the contact line, we provide the first 
theoretical explanation of a classical prediction due to 
Derjaguin and Levi: instability occurs when the outer, 
static meniscus approaches the shape
corresponding to a perfectly wetting fluid. 
\end{abstract}

\maketitle

The forced wetting or de-wetting of a solid is an important feature 
of many environmental and industrial flows. In typical
applications such as painting, coating, or oil recovery
it is crucial to know whether the solid will be covered 
by a macroscopic film or not. If a fiber is plunging into
a liquid bath to be coated (wetting case), the speed can be 
quite high (m/sec) \cite{SK03}, while maintaining a stationary contact line.
In the opposite case of withdrawal (de-wetting), a stationary contact line
is observed only for very low speeds, and a macroscopic 
film is deposited \cite{SP91,Q99} typically at a speed of 
only a few cm/sec. Yet no theoretical explanation for this instability 
exists, or of the fundamental difference between the two cases. 

It is well known \cite{HS71,DD74} that viscous forces become very large 
near a moving contact line, and are controlled only by some microscopic
cut-off $\lambda$, for example a slip length \cite{HS71,H83}. 
As a result of the interplay between viscous 
and surface tension forces, the interface is highly curved, and 
the contact line speed $U$ is properly measured by the capillary number
$Ca = U\eta/\gamma$, where $\eta$ is the viscosity of the fluid and 
$\gamma$ the surface tension between fluid and gas. Owing to this 
bending the interface angle measured at, say, $100 \mu m$ 
away from the contact line differs \cite{G85,K93,FJ91,MGD93} significantly
from the microscopic angle directly at the contact line. 

It was first proposed by Derjaguin and Levi \cite{DL64}, 
and later reiterated by others \cite{BR79}, 
that instability occurs if this dynamic interface angle 
reaches {\it zero}. Apart from the fact that there is no 
justification for this condition, it does not lead to a unique 
criterion since the angle depends on the position
where it is evaluated. However, it was noted experimentally \cite{SP91} 
that the interface profile at the critical speed corresponds to
a {\it static} meniscus with zero equilibrium contact angle,
except in the immediate neighborhood of the contact line. Another 
important set of experimental observations \cite{Q91} was performed with
fluids of different viscosities and equilibrium contact angles $\theta_e$ 
in a capillary tube: it was found that instability occurred if
$Ca/\theta_e^3$ exceeds a critical value, strongly pointing to 
a {\it hydrodynamic} mechanism for the instability. 

To explain the physical mechanism behind the instability, we note 
that the interface shape $h(x)$ near the contact line must have the 
form 
\begin{equation}  
\label{scal}  
h(x) = 3\lambda H(\xi), \quad \xi=x\theta_e/(3\lambda),
\end{equation}  
since the cut-off $\lambda$ is the only available length scale. The 
dependence on $\theta_e$ was put in for later convenience. As expected, 
the curvature $h''(0)=\theta_e^2/(3\lambda)H''(0)$ becomes very large at 
the contact line, since $\lambda$ is in the order of nanometers. 
As the curvature of the local solution has to match to a value of order unity 
in the static outer part of the profile, the usual boundary condition
for $H$ is one of vanishing curvature for large $\xi$. In the wetting
case this leads to an asymptotic solution first found
by Voinov \cite{V76} and an expression for the interface angle usually
referred to as ``Tanner's law'' \cite{G85}. 

However, it is a well-known but little appreciated fact \cite{DW97}
that Voinov's solution fails away from the contact line in the 
de-wetting situation. Instead, the local solution always 
retains a positive curvature, and will fail to match to an 
outer solution in the limit of small $\lambda$. At the critical 
speed, the necessary compromise between the inner solution near 
the contact line and the outer static solution has been pushed 
to the limit: from all possible inner solutions, the one with the 
{\it smallest} possible curvature is selected. The outer 
solution, on the other hand, realizes the solution with the {\it largest}
curvature, which happens to be the one corresponding to zero contact
angle, in agreement with the criterion of Derjaguin and Levi. Note that
this mechanism differs from the one proposed by de Gennes \cite{G86},
which is based entirely on properties of the {\it local} solution 
near the contact line. In a companion paper \cite{E03} we show
that de Gennes' local solution is inconsistent with a class of 
contact line models, such as the slip model considered here. 

\begin{figure}
\includegraphics[width=1\hsize]{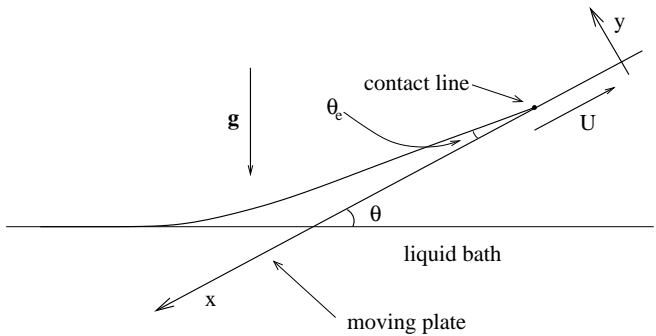}
\caption{\label{geom} 
 A schematic of the setup. At the contact line, h(0) = 0, and
the slope of the interface is $\theta_e$. The plate is being withdrawn
at an angle $\theta$. 
   }
\end{figure}

To lay down the theory behind the above argument, we choose a plate 
being withdrawn from a liquid bath (see Fig.\ref{geom}) as a test case, 
a geometry which was studied in detail in \cite{H01}. There it was found 
numerically that there exists a critical capillary number $Ca_{cr}$, 
above which no static solution exists, in agreement with the experimental
findings described above. If both $\theta$ 
and the equilibrium contact angle $\theta_e$ are small, one can 
treat the problem in the framework of lubrication theory, assuming 
a parabolic flow profile. This greatly simplifies the problem, 
but without altering its essential structure.

To relieve the corner singularity at the moving contact line, 
we allow the fluid to slide across the solid surface, following 
the Navier slip law \cite{H83} 
\begin{equation}  
\label{Navier}  
u|_{y=0}-U = \lambda\frac{\partial u}{\partial y}|_{y=0}
\end{equation}
at the plate, where $h(x)$ is the thickness of the fluid layer. 
The resulting lubrication equation is \cite{H01}
\begin{equation}  
\label{lub}  
\frac{3 Ca}{h^2 + 3\lambda h} = h''' - h' + \theta,
\end{equation}  
where we consistently used the small-angle approximation
$\tan(\theta)\approx\theta$. All lengths are scaled by the 
capillary length $\ell_c=\sqrt{\gamma/(\rho g)}$.

The left hand side of (\ref{lub}) corresponds to viscous forces, diverging 
as the contact line position $h(0)=0$ is approached, but weakened 
by the presence of slip. Close to the contact line, viscous forces 
are balanced by surface tension forces (first term on the right),
resulting in a highly curved interface near the contact line. 
The other two terms stem from gravity and only come into play
at greater distances. 
We also assume that the angle at the contact line $h'(0)=\theta_e$
is constant, which then has to be the equilibrium contact angle, 
in order to give the right result at vanishing speed. 
Note that we have used the simplest possible contact line 
model, and no special significance should be attached to our
choice. Since the instability mechanism is hydrodynamic, any
local contact line model can be incorporated into our description.
Far away from the interface the surface coincides with the 
liquid bath, so the third boundary condition is $h'(\infty)=\theta$.

To disentangle the local contact line behavior from the far-field 
meniscus, we transform equation (\ref{lub})
into similarity variables (\ref{scal}):
\begin{equation}  
\label{sim}  
\frac{\delta}{H^2 + H} = H''' + \mu^2(\theta/\theta_e - H')
\end{equation}  
with boundary conditions $H(0) = 0, H'(0)=1$, and 
$H'(\infty)=\theta/\theta_e$. 
The three parameters remaining in the problem are the rescaled 
capillary number $\delta=3Ca/\theta_e^3$, the rescaled cut-off 
$\mu=\lambda/\theta_e$, and the relative inclination angle
$\theta/\theta_e$.

Since we are interested in the limit of $\mu$ being very small, 
we will first look at the equation for $\mu=0$, as
appropriate close to the contact line. As mentioned above,
in the limit of large $\xi$ this equation behaves
very differently, depending on whether $\delta > 0$ or $\delta < 0$. 
This can be understood from the simplified equation 
$H''' = \delta/H^2$, which can be integrated completely \cite{BO78,DW97}.
Namely, if $\delta <0$ (wetting), there exists an asymptotic solution 
for $\xi\rightarrow\infty$ \cite{V76}, whose curvature 
{\it vanishes} at infinity. 

If on the other hand $\delta > 0$ (de-wetting), all solutions have 
a {\it finite} curvature at infinity. Solutions which obey our
boundary condition $H(0)=0$ can be written \cite{DW97} parametrically
as 
\begin{eqnarray}
\label{par}
\left.\begin{array}{l}
\xi = \frac{2^{1/3}\pi Ai(s)}{\beta(\alpha Ai(s) + \beta Bi(s))} \\    
H = \frac{\delta^{1/3}}{(\alpha Ai(s) + \beta Bi(s))^2}
                 \end{array}\right\}s\in [s_1\infty[,
\end{eqnarray}
where $Ai$ and $Bi$ are Airy functions. 
The limit $\xi\rightarrow 0$ corresponds to $s\rightarrow \infty$.
For large $\xi$, the curvature becomes \cite{DW97}
\begin{equation}  
\label{curv}  
\kappa_{\infty} = \left(\frac{2^{1/6}\beta}{\pi Ai(s_1)}\right)^2 > 0,
\end{equation}  
where $s_1$ solves the equation
\begin{equation}  
\label{s1}  
\alpha Ai(s_1)+\beta Bi(s_1)=0. 
\end{equation}  
Since the solution extends to
$s=\infty$, $s_1$ has to be the {\it largest} root of (\ref{s1}).

The constant $\beta$ can be determined by matching (\ref{par}) to the 
cutoff region near the contact line, where one finds 
$H'(\xi) \approx \left[3\delta\ln(2^{2/3}\beta^2/(\pi\xi)\right]^{1/3}$
in the limit of small $\xi$ \cite{DW97}. Comparing this to the first 
order expansion of the full equation (\ref{sim}) near the contact 
line \cite{H83,E03}, 
\begin{equation}  
\label{exp}  
H'(\xi)=1-\delta(1 + \ln(\xi)),
\end{equation}  
we find $\beta^2=\pi\exp(1/(3\delta))/2^{2/3} + O(\delta)$. 
The matching described here 
was investigated in greater detail for the case $\delta < 0$ in
\cite{E03}. We found that higher order corrections in $\delta$ 
were surprisingly weak, and depended only very little on the type
of cut-off used at the contact line. Thus we are confident that 
the same holds true in the present case. 

At the critical capillary number $Ca_{cr}$ the only remaining parameter,
which is $\alpha$, can be determined from the condition that 
$\kappa_{\infty}$ should be {\it minimal}, in other words 
$Ai(s_1)$ must be maximal among solutions of (\ref{s1}).
By choosing $\alpha=\alpha_{cr}\equiv-\beta Bi(s_{max})/Ai(s_{max})$
we can in fact ensure that $Ai$ assumes its global maximum 
$0.53566\dots$, which occurs for $s=s_{max}=-1.0188\dots$. 
In summary, we find then for the critical solution 
that minimizes the curvature 
\begin{equation}  
\label{ccr}  
\kappa^{cr}_{\infty} = \frac{\delta^{1/3}\exp[-1/(3\delta)]}
{2^{1/3}\pi (Ai(s_{max}))^2},
\end{equation}  
which is now given exclusively in terms of the rescaled capillary
number $\delta$. 

The solution (\ref{par}) has to be matched to an outer solution of
(\ref{lub}), valid away from the contact line. In the spirit of the 
classical technique \cite{H83} employed for the spreading drops, 
this is a {\it static} solution of (\ref{lub}), but with an {\it apparent}
contact angle different from $H'(0)=1$. 
The general solution of (\ref{sim}) with $\delta=0$ gives 
\begin{equation}  
\label{static}  
H'(\xi) = \theta/\theta_e - \theta_0\ e^{-\mu\xi},
\end{equation}  
so the curvature at the plate is $H''(0) = \mu\theta_0$. The {\it largest} 
curvature for which (\ref{static}) still makes sense is 
$\mu\theta/\theta_e$, at which point the apparent contact angle has 
reached zero. 

The two solutions have to be matched together, such that the 
curvature of (\ref{par}) for large arguments agrees with that of 
(\ref{static}) for small arguments. At large capillary number this
will no longer be possible, since $\kappa_{\infty}^{cr}$ becomes
larger than $\mu\theta/\theta_e$. At the critical capillary 
number $Ca_{cr}$, (\ref{ccr}) thus has to be equated with 
$\mu\theta/\theta_e$, giving 
\begin{equation}  
\label{dcr}  
\delta_{cr}^{1/3}\exp[-1/(3\delta_{cr})] = 
{2^{1/3}\pi (Ai(s_{max}))^2}\mu\theta/\theta_e
\end{equation}  
for the critical rescaled capillary number 
$\delta_{cr} = 3Ca_{cr}/\theta_e^3$. This equation completely 
determines the stability boundary found numerically in \cite{H01},
as function of all parameters. It also implies that for a given 
geometry, instability occurs if the parameter 
$\delta_{cr}=3Ca/\theta_e^3$ exceeds a critical value
(up to very small logarithmic corrections), in agreement
with experiment \cite{Q91}. 

For our procedure to be consistent, though, we need to make sure 
that the two solutions (\ref{par}) and (\ref{static}) 
have sufficient overlap to be matched. The inner and outer 
solutions at $Ca_{cr}$ are, in summary, 
\begin{eqnarray*}
\label{sum}
&&H'^{cr}_{inner} = \delta^{1/3}f(\xi\beta^2) \\
&&H'^{cr}_{outer} = (\theta/\theta_e)(1 - e^{-\mu\xi}),
\end{eqnarray*}
where $f$ is a universal function given by (\ref{par}) with 
$\alpha = \alpha_{cr}$. 
Thus if $\beta^2\gg\mu$ the large-$\xi$ limit of $H'^{cr}_{inner}$
overlaps with the small-$\xi$ limit of $H'^{cr}_{outer}$, which
translates into $\theta/\theta_e\gg\delta^{1/3}$. 

\begin{figure}
\includegraphics[width=1\hsize]{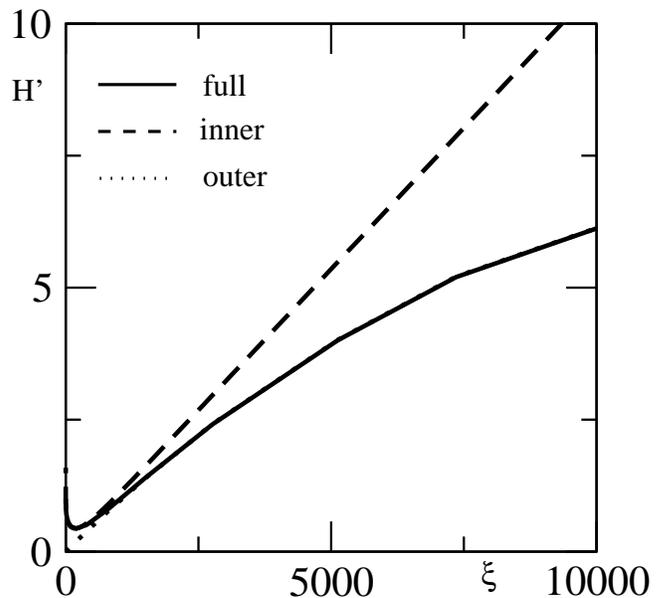}
\caption{\label{compare} 
A comparison of the full solution at the critical capillary number with
the inner and outer solutions. We plot the rescaled slope, so $H'(0)=1$
for the full solution, and $H'(0)=0$ for the outer solution, consistent
with the condition by Derjaguin and Levi. The parameters are $\mu=10^{-4}$ 
and $\theta/\theta_e=10$, and thus $\delta_{cr}=0.0572$ from (\ref{dcr}).  
   }
\end{figure}

\begin{figure}
\includegraphics[width=1\hsize]{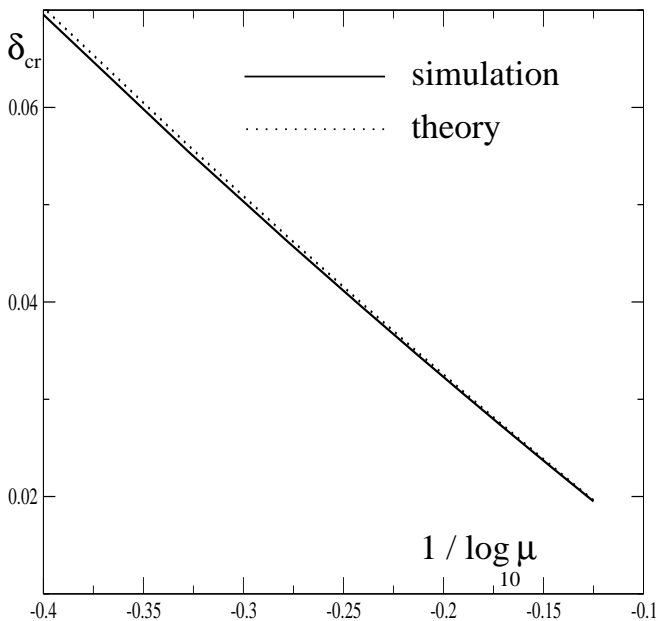}\caption{\label{theory} 
A comparison of $\delta_{cr}$ as determined by numerical integration
of equation (\ref{sim}) for $\theta/\theta_{cr} = 1$, and theory,
as summarized by equation (\ref{dcr}).
   }
\end{figure}

In Fig. \ref{compare} we show the result of a numerical integration
of (\ref{sim}) for $\theta/\theta_{cr} = 10$ at the critical capillary
number. The interface slope follows the static solution perfectly 
almost up to the contact line, where it has to turn over, while
the static solution extrapolates to zero. Coming from the interior, 
(\ref{par}) describes the solution equally well up to the turning 
point. Note that there are no adjustable parameters in Fig. \ref{compare},
we simply took the inner and outer solutions in the critical case. 
In fact, even for $\theta/\theta_{cr} = 1$, when there is not yet
much overlap between the two solutions, equation (\ref{dcr}) already
works extremely well in describing the loss of the stationary 
solution, as shown in Fig. \ref{theory}. Again, no parameter was
adjusted to achieve this comparison. 

Beyond the wet-dry transition studied here, in \cite{H01} 
two more states have been described, characterized by thin and thick wetting
films, respectively. The stability of these solutions, and possible 
hysteresis phenomena remain to be investigated. It is important to note
that our approach is not limited to the moving plate geometry 
studied here, nor is it restricted to a specific contact line model,
since it is based entirely on hydrodynamic arguments {\it away}
from the contact line. For example, if van-der-Waals forces are
dominant near the contact line \cite{GHL90}, the parameter $\mu$ 
in (\ref{dcr}) simply needs to be replaced by 
$\mu_{vdw}=\sqrt{A/(6\pi\gamma)}/(2\theta_e^2)$, where $A$ is the 
Hamaker constant. 

To generalize to a different geometry, one has to replace
(\ref{static}) by the appropriate static solution for the problem
at hand. This is done almost trivially for the case of a vertical
wall \cite{LL84}, and easily extended \cite{H01} to the flow in a
narrow capillary, to be able to compare directly to experiments 
\cite{Q91}. Hocking \cite{H01} found that the present slip theory
correctly predicts $Ca_{cr}/\theta_e^3$ to be a constant, but
overestimates this constant by a factor of two, if reasonable
values for the slip length $\lambda$ are assumed. Reasons for 
this discrepancy could be both the considerable contact angle 
hysteresis in the experiment \cite{Q91}, and failure of the simple
contact line model used in our calculation. A new set of experiments, 
using the plate geometry, is being planned to clear up these 
questions \cite{Q03}. 

Another important generalization is to higher dimensional
problems, in which the contact line does not remain straight. An
instability toward inclined contact lines was observed in \cite{BR79}, 
as well as in recent experiments with
drops running down an inclined plane \cite{PFL01}. To explain the
characteristic inclination angle of such a contact line, one needs
to identify a {\it characteristic speed} of de-wetting \cite{BR79,SLW02},
which our present approach effectively provides. 

\acknowledgments
Special thanks are due to Petr Braun for numerous insightful
discussions.

\end{document}